\documentclass{PoS}
\usepackage{amsmath}
\usepackage{slashed}
\usepackage{amsfonts,amsmath,amssymb,bm,graphicx,cite,makeidx,multicol,color}

\title{Electromagnetic properties of neutrinos}

\ShortTitle{Electromagnetic neutrinos}

\author{\speaker{Alexander Studenikin}\\
        Department of Theoretical Physics, Moscow State University, 119992 Moscow, Russia\\
        Joint Institute for Nuclear Research, 141980 Dubna, Moscow Region, Russia\\
        E-mail: \email{studenik@srd.sinp.msu.ru}}

\abstract{After a brief introduction to neutrino electromagnetic properties
we focus on the most important constraints on neutrino magnetic moments,
charge radii and millicharges from the terrestrial experiments and astrophysical
considerations. The promising new possibilities for constraining neutrino
electromagnetic properties in future experiments are also discussed.}

\FullConference{
European Physical Society Conference on High Energy Physics - EPS-HEP2019 -\\
			10-17 July, 2019\\
			Ghent, Belgium}

\begin{document}

%\section{Neutrino electromagnetic form factors}
%\section{Neutrino magnetic moment}
%\section{Neutrino electric moment}
\section{Introduction}
There is no doubt \cite{Giunti:2014ixa,Studenikin:2008bd,Studenikin:2018vnp} that neutrino electromagnetic properties open a window to {\it new physics}.
The most general form \cite{Giunti:2014ixa} of a neutrino electromagnetic vertex function
\begin{equation}
\Lambda_{\mu}^{ij}(q) =  \left( \gamma_{\mu} - q_{\mu}
\slashed{q}/q^{2} \right) \left[ f_{Q}^{ij}(q^{2}) + f_{A}^{ij}(q^{2})
q^{2} \gamma_{5} \right] \nonumber
 - i \sigma_{\mu\nu} q^{\nu} \left[ f_{M}^{ij}(q^{2}) +
i f_{E}^{ij}(q^{2}) \gamma_{5} \right] ,
\end{equation}
where $\Lambda_{\mu}(q)$ and form factors $f_{Q,A,M,E}(q^2)$ are $3\times 3$ matrices in  the space of massive neutrinos, in the case of coupling with a real photon ($q^2=0$) provides four sets of neutrino electromagnetic characteristics: 1) the dipole magnetic moments $\mu_{ij}=f_{M}^{ij}(0)$,
2) the dipole electric moments $\epsilon_{ij}=f_{E}^{ij}(0)$, 3) the millicharges $q_{ij}=f_{Q}^{ij}(0)$ and
4) the anapole moments $a_{ij}=f_{A}^{ij}(0)$.

So far, there are no indications in favor of nonzero electromagnetic properties of neutrinos from either data from laboratory experiments with neutrino fluxes from ground-based sources or from astrophysics observations.  However, the study of the electromagnetic properties of neutrinos attracts considerable attention.

\section{Neutrino magnetic and electric dipole moments}

The most well understood and studied are the dipole magnetic and electric moments.
In a minimal extension of the Standard Model  the diagonal magnetic  moment of a Dirac neutrino is given \cite{Fujikawa:1980yx} by

\begin{equation}
\mu^{D}_{ii}
  = \frac{3e G_F m_{i}}{8\sqrt {2} \pi ^2}\approx 3.2\times 10^{-19}
  \Big(\frac{m_i}{1 \ \mathrm{eV} }\Big) \mu_{B},
  \end{equation}
$\mu_B$ is the Bohr magneton. The Majorana neutrinos can have  only transition
(off-diagonal) magnetic
moments  $\mu^{M}_{i\neq j}$. However, in the flavour basis
the diagonal magnetic and electric moments can be nonzero even in the case of the
Majorana neutrinos (for instance, when the diagonal magnetic moments for the neutrino
mass states are not equal, $\mu_{ii}\neq\mu_{jj}$).

The most stringent constraints on the effective neutrino
magnetic moment are obtained with the reactor antineutrinos
(GEMMA Collaboration \cite{GEMMA:2012})
\begin{equation}\label{GEMMA}
\mu_{\nu} < 2.9 \times 10^{-11} \mu_{B},
\end{equation}
and solar neutrinos (Borexino Collaboration \cite{Borexino:2017fbd})
\begin{equation}\label{Borexino}
{\mu}_{\nu_e}\leq 2.8 \times
10^{-11} \mu _B.
\end{equation}

It should be noted, that in general in scattering experiments the neutrino
is created at some distance from the detector as a flavor neutrino, which is a
superposition of massive neutrinos. Therefore, the magnetic
and electric moments that is measured in these experiment is not that of a
massive neutrino, but it is effective moments which takes into account neutrino mixing and the oscillations
during the propagation between source and detector \cite{Grimus:1997aa, Beacom:1999wx}.
For the recent and detailed study of the neutrino electromagnetic characteristics
dependence on neutrino mixing see \cite{Kouzakov:2017hbc}.

An astrophysical bound (for both
Dirac and Majorana neutrinos) is provided
\cite{Raffelt-Clusters:90, Viaux-clusterM5:2013, Arceo-Diaz-clust-omega:2015}
by observations of the properties of globular cluster stars
\begin{equation}
\Big( \sum _{i,j}\left| \mu_{ij}\right| ^2\Big) ^{1/2}\leq (2.2{-}2.6) \times
10^{-12} \mu _B.\end{equation}
A general and termed model-independent upper bound on the Dirac neutrino
magnetic moment, that can be generated by an effective theory beyond
a minimal extension of the Standard Model, has been derived in
\cite{Bell:2005kz}: $\mu_{\nu}\leq
10^{-14}\mu_B$. The corresponding limit for transition moments of Majorana neutrinos is much weaker \cite{Bell:2006wi}.

%\section{Neutrino electric moment}

\ \ \ \ \ \ In the theoretical framework with $CP$ violation a neutrino
can have nonzero electric moments $\epsilon_{ij}$. In the laboratory neutrino
scattering experiments for searching $\mu_{\nu}$ (for instance, in the GEMMA experiment)
the electric moment $\epsilon_{ij}$ contributions interfere with
those due to $\mu_{ij}$. Thus, these kind of experiments also provide constraints
on $\epsilon_{ij}$. The astrophysical bounds on $\mu_{ij}$
are also applicable for constraining $\epsilon_{ij}$ (see \cite{Raffelt-Clusters:90, Viaux-clusterM5:2013, Arceo-Diaz-clust-omega:2015} and \cite{Raffelt:2000kp}).

In what follows below we give a fast flash on less know neutrino electromagnetic characteristics, namely
 on the neutrino millicharge, charge radius and anapole moment and give some comments on the future prospects of neutrino electromagnetic properties.
\section{Neutrino electric millicharge}
There are extensions of the Standard Model that allow for nonzero
neutrino electric millicharges. This option can be provided by
not excluded experimentally possibilities for hypercharhge dequantization or
another {\it new physics} related with an additional $U(1)$ symmetry
peculiar for extended theoretical frameworks. Neutrino millicharges
are strongly constrained on the level $q_{\nu}\sim 10^{-21} e_0$
($e_0$ is the value of an electron charge) from neutrality of the hydrogen atom.

 A nonzero neutrino millicharge $q_{\nu}$ would contribute to the neutrino electron scattering in the terrestrial experiments. Therefore, it is possible to get bounds on $q_{\nu}$ in the reactor antineutrino
 experiments. The most stringent reactor antineutrino constraint
 \begin{equation}
 q_{\nu}\leq 1.5 \times 10^{-11} e_0
 \end{equation}
 is obtained in \cite{Studenikin:2013my} (see also \cite{PDG2016})
 with use of the GEMMA experimental data \cite{GEMMA:2012}.

A neutrino millicharge might have specific phenomenological consequences
in astrophysics because of new electromagnetic processes are opened
due to a nonzero charge (see \cite{Giunti:2014ixa,Raffelt:1996wa}). Following this line, the most stringent astrophysical constraint on neutrino millicharges
\begin{equation}
q_{\nu}\leq 1.3 \times 10^{-19} e_0
\end{equation}
 was obtained in \cite{Studenikin:2012vi}. This bound
follows from the impact of the {\it neutrino star turning} mechanism ($ST\nu$) \cite{Studenikin:2012vi} that can be charged as a {\it new physics} phenomenon end up with a pulsar rotation frequency
shift engendered by the motion of escaping from the
star neutrinos along curved trajectories due to millicharge interaction with a constant
magnetic field of the star.

\section{Neutrino charge radius and anapole moment}
Even if a neutrino millicharge is vanishing, the electric form factor
$f^{ij}_{Q}(q^{2})$ can still contain nontrivial information about
neutrino electromagnetic properties. The corresponding electromagnetic characteristics is
determined by the derivative of $f^{ij}_{Q}(q^{2})$ over $q^{2}$  at
$q^{2}=0$ and is termed neutrino charge radius,
%\begin{equation}
%\label{nu_cha_rad_1}
$\langle{r}_{ij}^{2}\rangle
=-
6
%\left.
\frac{df^{ij}_{Q}(q^{2})}{dq^{2}} \
_{\mid _ {q^{2}=0}}
$ (see \cite{Giunti:2014ixa} for the detailed discussions).
%\end{equation}
Note that for a massless neutrino the neutrino charge radius is the only
electromagnetic characteristic that can have nonzero value. In the Standard Model
the neutrino charge radius and the anapole moment are not defined separately,
and there is a relation between these two values: $a = - \frac{\langle{r}^{2}\rangle}{6}$.

A neutrino charge radius contributes to the neutrino scattering cross section on electrons and thus
can be constrained by the corresponding laboratory experiments \cite{Bernabeu:2004jr}.
In all but one previous studies it was claimed
 that the effect of the neutrino
charge radius can be included just as a shift of the vector coupling constant $g_V$
in the weak
contribution to the cross section.
However, as it has been recently demonstrated in \cite{Kouzakov:2017hbc} within the direct calculations of
the elastic neutrino-electron scattering cross section accounting for all possible neutrino electromagnetic characteristics
and neutrino mixing, this is not the fact. The neutrino charge radius dependence of the cross section
is more complicated and there are, in particular, the dependence on the interference terms of the type
$g_{V}\langle{r}_{ij}^{2}\rangle$ and also on the neutrino mixing.

\section{Conclusions and future prospects}
 The foreseen progress in constraining neutrino electromagnetic characteristics is related, first of all, with the expected new results from the GEMMA experiment measurements of the reactor antineutrino cross section on electrons at Kalinin Power Plant. The new set of data is expected to arrive next year. The electron energy threshold will be as low as $350 \ eV$ ( or even lower, up to $\sim 200 \ eV$). This will provide possibility to test the neutrino magnetic moment on the level of $\mu_\nu \sim 0.9 \times 10^{-12} \mu_B$ and also to test the millicharge on the level of $q_{\nu} \sim 1.8 \times 10^{-13} e_0$ \cite{Studenikin:2013my}.

The current constraints on the flavour neutrino charge radius $\langle{r}_{e,\mu,\tau}^{2}\rangle\leq 10^{-32} - 10^{-31} \ cm ^2$
from the scattering experiments differ only by 1 to 2
orders of magnitude from the values $\langle{r}_{e,\mu,\tau}^{2}\rangle\leq 10^{-33} \ cm ^2$ calculated within the minimally extended Standard Model with right-handed neutrinos
\cite{Bernabeu:2004jr}. This indicates that the minimally extended Standard Model neutrino charge radii could be experimentally tested in the near future.

Note that there is a need to re-estimate experimental constraints on
$\langle{r}_{e,\mu,\tau}^{2}\rangle$  from the scattering experiments following
new derivation of the cross section \cite{Kouzakov:2017hbc} that properly accounts for the interference of the weak and charge radius electromagnetic interactions and also for the neutrino mixing.

Recently constraints on  charged radii  have been obtained
\cite{Caddedu:2018prd} from the analysis of the data on coherent
elastic neutrino-nucleus scattering obtained in the COHERENT experiment
\cite{Akimov:2017ade,Akimov:2018vzs}. In addition to the customary diagonal
charge radii $\langle{r}_{e,\mu,\tau}^{2}\rangle$
also the neutrino transition (off-diagonal) charge radii have been constrained
in \cite{Caddedu:2018prd} for the first time:
$$
\left(|\langle r_{\nu_{e\mu}}^2\rangle|,|\langle r_{\nu_{e\tau}}^2\rangle|,|\langle r_{\nu_{\mu\tau}}^2\rangle|\right)
< (22,38,27)\times10^{-32}~{\rm cm}^2.
$$

Quite recently the potential of current and next generation of coherent elastic
neutrino-nucleus scattering experiments in probing neutrino electromagnetic
interactions has been explored \cite{Miranda:2019wdy}. In particular, the present stage
and the next phase of the COHERENT experiment, as well as several other reactor
experiments  sensitivities to the Majorana neutrino transition magnetic moments is estimated.
The conclusion is that future experiments with low-threshold capabilities can
improve current limits on transition magnetic moments obtained from Borexino data.

For the future progress in studying (or constraining) neutrino electromagnetic properties
a rather promising claim was made in  \cite{deGouvea:2012hg,deGouvea:2013zp}. It was shown that
even tine values of the Majorana neutrino transition moments
would probably be tested in future high-precision experiments with the astrophysical neutrinos.  In particular,
observations of supernova fluxes  in the JUNO  experiment (see
\cite{An:2015jdp,Giunti:2015gga,Lu:2016ipr})
may reveal the effect of  collective  spin-flavour oscillations  due to the Majorana neutrino transition moment $\mu^{M}_\nu \sim 10^{-21}\mu_B$. There are indeed other new possibilities for neutrino
magnetic moment visualization in extreme astrophysical environments
considered recently \cite{Grigoriev:2017wff,Kurashvili:2017zab}.

In the most recent paper \cite{Cadeddu:2019qmv} we have proposed an experimental setup to observe coherent elastic neutrino-atom scattering using electron antineutrinos from tritium decay and a liquid helium target. In this scattering process with the whole atom, that has not beeen observed so far, the electrons tend to screen the weak charge of the nucleus as seen by the electron antineutrino probe.
 Finally, we study the sensitivity of this apparatus to a possible electron
 neutrino magnetic moment and we find that it is possible
 to set an upper limit of about
 \begin{equation}
\mu_{\nu} < 7 \times 10^{-13} \mu_{B},
\end{equation}
at 90 \%  C.L.,  that is more than one order of magnitude smaller than
the current experimental limits (\ref{GEMMA}) and (\ref{Borexino}).


\begin{thebibliography}{99}
\bibitem{Giunti:2014ixa}
  C.~Giunti and A.~Studenikin,
  \emph {Neutrino electromagnetic interactions: a window to {\it new physics},
  Rev.\ Mod.\ Phys.\ } {\bf 87} (2015) 531.
  %%CITATION = doi:10.1103/RevModPhys.87.531;%%
  %48 citations counted in INSPIRE as of 21 Nov 2016
\bibitem{Studenikin:2008bd}
  A.~Studenikin,
  \emph {Neutrino magnetic moment: a window to {\it new physics},
  Nucl.\ Phys.\ Proc.\ Suppl.\ } {\bf 188} (2009) 220.
  %doi:10.1016/j.nuclphysbps.2009.02.053
  %[arXiv:0812.4716 [hep-ph]].
  %%CITATION = doi:10.1016/j.nuclphysbps.2009.02.053;%%
  %22 citations counted in INSPIRE as of 10 Oct 2017
%\cite{Studenikin:2018vnp}
\bibitem{Studenikin:2018vnp}
  A.~Studenikin,
  \emph {Neutrino electromagnetic properties: a window to new physics - II,
  PoS EPS-HEP2017} (2017) 137
  %doi:10.22323/1.314.0137
  %[arXiv:1801.08887 [hep-ph]].
  %%CITATION = doi:10.22323/1.314.0137;%%
  %5 citations counted in INSPIRE as of 09 Nov 2018
\bibitem{Fujikawa:1980yx}
  K.~Fujikawa and R.~Shrock,
  \emph{The magnetic moment of a massive neutrino and neutrino spin rotation,
  Phys.\ Rev.\ Lett.\ }  {\bf 45} (1980) 963.
%\cite{Raffelt:2000kp}
\bibitem{GEMMA:2012}
A.~G. Beda, V.~B. Brudanin, V.~G. Egorov
%D.~V. Medvedev, V.~S. Pogosov, M.~V.   Shirchenko
et~al.,
  \emph{{The results of search for the neutrino magnetic
  moment in GEMMA experiment,
  Adv. High Energy Phys.} {\bf 2012}
  (2012) 350150}.
\bibitem{Borexino:2017fbd}
  M.~Agostini {\it et al.} [Borexino Collaboration],
  \emph{Limiting neutrino magnetic moments with Borexino Phase-II solar neutrino data},
  arXiv:1707.09355 [hep-ex].
  %%CITATION = ARXIV:1707.09355;%%
  %1 citations counted in INSPIRE as of 19 Sep 2017
%\cite{Grimus:1997aa}
\bibitem{Grimus:1997aa}
  W.~Grimus and P.~Stockinger,
  %``Effects of neutrino oscillations and neutrino magnetic moments on elastic neutrino - electron scattering,''
  Phys.\ Rev.\ D {\bf 57} (1998) 1762
  doi:10.1103/PhysRevD.57.1762
  [hep-ph/9708279].
  %%CITATION = doi:10.1103/PhysRevD.57.1762;%%
  %17 citations counted in INSPIRE as of 01 Nov 2019
%\cite{Beacom:1999wx}
\bibitem{Beacom:1999wx}
  J.~F.~Beacom and P.~Vogel,
  %``Neutrino magnetic moments, flavor mixing, and the Super-Kamiokande solar data,''
  Phys.\ Rev.\ Lett.\  {\bf 83} (1999) 5222
  doi:10.1103/PhysRevLett.83.5222
  [hep-ph/9907383].
  %%CITATION = doi:10.1103/PhysRevLett.83.5222;%%
  %159 citations counted in INSPIRE as of 01 Nov 2019
\bibitem{Kouzakov:2017hbc}
  K.~Kouzakov and A.~Studenikin,
  \emph {Electromagnetic properties of massive neutrinos in low-energy elastic neutrino-electron scattering,
  Phys.\ Rev.\ D } {\bf 95} (2017) 055013.
  %doi:10.1103/PhysRevD.95.055013
  %[arXiv:1703.00401 [hep-ph]].
  %%CITATION = doi:10.1103/PhysRevD.95.05501
\bibitem{Raffelt-Clusters:90}
G.~G. Raffelt,
\emph{New bound on neutrino dipole moments from
  globular-cluster stars,
 Phys. Rev. Lett.} {\bf 64} (1990) 2856.
  \bibitem{Viaux-clusterM5:2013}
N.~Viaux, M.~Catelan, P.~B. Stetson, G.~G. Raffelt
%J.~Redondo, A.~A.~R.   Valcarce
et~al.,
  \emph{Particle-physics constraints from the globular cluster
  {M5}: neutrino dipole moments,
  Astron. \& Astrophys.} {\bf 558}
  (2013) A12.
\bibitem{Arceo-Diaz-clust-omega:2015}
S.~Arceo-D\'{i}az, K.-P. Schr\"{o}der, K.~Zuber and D.~Jack,
\emph{Constraint
  on the magnetic dipole moment of neutrinos by the tip-RGB luminosity in
  $\omega$-Centauri,
  %\emph
  Astropart. Phys.} {\bf 70} (2015) 1.
%\cite{Bell:2005kz}
\bibitem{Bell:2005kz}
  N.~F.~Bell, V.~Cirigliano, M.~J.~Ramsey-Musolf et al
  % P.~Vogel and M.~B.~Wise,
  \emph {How magnetic is the Dirac neutrino?,
  Phys.\ Rev.\ Lett.\ } {\bf 95} (2005) 151802.
  %doi:10.1103/PhysRevLett.95.151802
  %[hep-ph/0504134].
  %%CITATION = doi:10.1103/PhysRevLett.95.151802;%%
  %80 citations counted in INSPIRE as of 11 Oct 2017
\bibitem{Bell:2006wi}
  N.~F.~Bell, M.~Gorchtein, M.~J.~Ramsey-Musolf, P.~Vogel and P.~Wang,
  \emph {Model independent bounds on magnetic moments of Majorana neutrinos,
  Phys.\ Lett.\  B }{\bf 642} (2006) 377.
  %doi:10.1016/j.physletb.2006.09.055
  %[hep-ph/0606248].
  %%CITATION = doi:10.1016/j.physletb.2006.09.055;%%
  %56 citations counted in INSPIRE as of 11 Oct 2017
\bibitem{Raffelt:2000kp}
  G.~G.~Raffelt,
  \emph {Astrophysics probes of particle physics,
  Phys.\ Rept.\ } {\bf 333} (2000) 593.
  %doi:10.1016/S0370-1573(00)00039-9
  %%CITATION = doi:10.1016/S0370-1573(00)00039-9;%%
  %39 citations counted in INSPIRE as of 11 Oct 2017
\bibitem{Studenikin:2013my}
  A.~Studenikin,
  \emph {New bounds on neutrino electric millicharge from limits on neutrino magnetic moment,
  Europhys.Lett. } {\bf 107} (2014) 21001.
   %Erratum: [EPL {\bf 107} (2014) no.3,  39901]
  %doi:10.1209/0295-5075/107/21001, 10.1209/0295-5075/107/39901
  %[arXiv:1302.1168 [hep-ph]].
  %%CITATION = doi:10.1209/0295-5075/107/21001, 10.1209/0295-5075/107/39901;%%
  %15 citations counted in INSPIRE as of 15 Oct 2017
%\cite{Tanabashi:2018oca}
  \bibitem{PDG2016} C. Patrignani \emph{et al.}
  (Particle Data Group), \emph{Chin. Phys. C}
  \textbf{40} (2016) 100001; M. Tanabashi \emph{et al.}
  (Particle Data Group), \emph{Phys. Rev. D} \textbf{98}
  (2018) 030001 and 2019 update.
%\cite{Raffelt:1996wa}
\bibitem{Raffelt:1996wa}
  G.~G.~Raffelt,
  \emph {Stars as laboratories for fundamental physics : The astrophysics of neutrinos, axions, and other weakly interacting particles},
  Chicago, USA: Univ. Pr. (1996) 664 p.
  %108 citations counted in INSPIRE as of 15 Nov 2018
\bibitem{Studenikin:2012vi}
  A.~Studenikin and I.~Tokarev,
  \emph {Millicharged neutrino with anomalous magnetic moment in rotating magnetized matter,
  Nucl.\ Phys.\ B }{\bf 884} (2014) 396.
  %doi:10.1016/j.nuclphysb.2014.04.026
  %[arXiv:1209.3245 [hep-ph]].
  %%CITATION = doi:10.1016/j.nuclphysb.2014.04.026;%%
  %17 citations counted in INSPIRE as of 13 Oct 2017
%\cite{Bernabeu:2004jr}
\bibitem{Bernabeu:2004jr}
  J.~Bernabeu, J.~Papavassiliou and D.~Binosi,
  \emph {The neutrino charge radius in the presence of fermion masses,
  Nucl.\ Phys.\ B }{\bf 716} (2005) 352.
  %doi:10.1016/j.nuclphysb.2005.02.039
  %[hep-ph/0405288].
  %%CITATION = doi:10.1016/j.nuclphysb.2005.02.039;%%
  %14 citations counted in INSPIRE as of 16 Oct 2017
%\cite{Kouzakov:2017hbc}

\bibitem{Caddedu:2018prd} M. Caddedu, C. Giunti, K. A. Kouzakov,
Y.-F. Li, A. I. Studenikin and Y. Y. Zhang, \emph{%Neutrino charge radii from
COHERENT elastic neutrino-nucleus scattering,
Phys. Rev. D} \textbf{98} (2018) 113010,
  arXiv:1810.05606 [hep-ph].
  %%CITATION = ARXIV:1810.05606;%%
%\cite{Akimov:2017ade}
\bibitem{Akimov:2017ade}
  D.~Akimov {\it et al.} [COHERENT Collaboration],
   \emph {Observation of Coherent Elastic Neutrino-Nucleus Scattering,
  Science} {\bf 357} (2017) no.6356,  1123.
  %doi:10.1126/science.aao0990
  %[arXiv:1708.01294 [nucl-ex]].
  %%CITATION = doi:10.1126/science.aao0990;%%
  %87 citations counted in INSPIRE as of 14 Nov 2018%\cite{Akimov:2018vzs}
\bibitem{Akimov:2018vzs}
  D.~Akimov {\it et al.} [COHERENT Collaboration],
   \emph {COHERENT Collaboration data release from the first observation of coherent elastic neutrino-nucleus scattering,}
  %doi:10.5281/zenodo.1228631
  arXiv:1804.09459 [nucl-ex].
  %%CITATION = doi:10.5281/zenodo.1228631;%%
  %6 citations counted in INSPIRE as of 14 Nov 2018
%\cite{Miranda:2019wdy}
\bibitem{Miranda:2019wdy}
  O.~G.~Miranda, D.~K.~Papoulias, M.~Tórtola and J.~W.~F.~Valle,
  %``Probing neutrino transition magnetic moments with coherent elastic neutrino-nucleus scattering,''
  JHEP {\bf 1907} (2019) 103
  doi:10.1007/JHEP07(2019)103
  [arXiv:1905.03750 [hep-ph]].
  %%CITATION = doi:10.1007/JHEP07(2019)103;%%
  %10 citations counted in INSPIRE as of 01 Nov 2019

  \bibitem{deGouvea:2012hg} A. de Gouvea and S. Shalgar,
\emph{Effect of transition magnetic moments on collective supernova neutrino oscillations,
JCAP} \textbf{1210} (2012) 027.
%
\bibitem{deGouvea:2013zp} A. de Gouvea and S. Shalgar,
\emph{Transition magnetic moments and collective neutrino oscillations:
Three-flavor effects and detectability,
JCAP} \textbf{1304} (2013) 018.
%
\bibitem{An:2015jdp} F. An \emph{et al.} [JUNO Collaboration],
\emph{Neutrino physics with JUNO,
J. Phys. G} \textbf{43} (2016) 030401.
%
%\cite{Giunti:2015gga}
\bibitem{Giunti:2015gga}
  C.~ Giunti, K.~Kouzakov, Y.~F.~Li, A.~Lokhov, A.~Studenikin, S.~Zhou,
  \emph {Electromagnetic neutrinos in laboratory experiments and astrophysics,
  Annalen Phys.\ } {\bf 528} (2016) 198
  %doi:10.1002/andp.201500211
  %[arXiv:1506.05387 [hep-ph]].
  %%CITATION = doi:10.1002/andp.201500211;%%
  %17 citations counted in INSPIRE as of 16 Oct 2017
%%\cite{Lu:2016ipr}
\bibitem{Lu:2016ipr}
J.S. Lu, Y.-F. Li and S. Zhou,
\emph{Getting the most from the detection of Galactic supernova neutrinos in future large liquid-scintillator detectors,
Phys. Rev. D} {\bf 94} (2016) 023006.
%\cite{Grigoriev:2017wff}
\bibitem{Grigoriev:2017wff}
  A.~Grigoriev, A.~Lokhov, A.~Studenikin and A.~Ternov,
  \emph {Spin light of neutrino in astrophysical environments,
  JCAP} {\bf 1711} (2017) no.11,  024
  %doi:10.1088/1475-7516/2017/11/024
  %[arXiv:1705.07481 [hep-ph]].
  %%CITATION = doi:10.1088/1475-7516/2017/11/024;%%
  %3 citations counted in INSPIRE as of 06 Nov 2018
  %\cite{Kurashvili:2017zab}
\bibitem{Kurashvili:2017zab}
  P.~Kurashvili, K.~A.~Kouzakov, L.~Chotorlishvili and A.~I.~Studenikin,
  \emph {Spin-flavor oscillations of ultrahigh-energy cosmic neutrinos in interstellar space: The role of neutrino magnetic moments,
  Phys.\ Rev.\ D} {\bf 96} (2017) 103017
  %doi:10.1103/PhysRevD.96.103017
  %[arXiv:1711.04303 [hep-ph]].
  %%CITATION = doi:10.1103/PhysRevD.96.103017;%%
  %3 citations counted in INSPIRE as of 06 Nov 2018%\cite{Cadeddu:2018dux}

%\cite{Cadeddu:2019qmv}
\bibitem{Cadeddu:2019qmv}
  M.~Cadeddu, F.~Dordei, C.~Giunti, K.~A.~Kouzakov, E.~Picciau and A.~I.~Studenikin,
  Potentialities of a low-energy detector based on $^4$He evaporation to
  observe atomic effects in coherent neutrino scattering and physics perspectives,
  \emph {Phys. Rev. D} \textbf {100} (2019) 073014, arXiv:1907.03302 [hep-ph].
  %%CITATION = ARXIV:1907.03302;%%
  %2 citations counted in INSPIRE as of 31 Oct 2019

\end{thebibliography}
\end{document}